\documentclass[usenatbib]{mnras}
\usepackage{lscape}
\usepackage{amsmath,amssymb}
\usepackage{mathrsfs}	
\usepackage{multicol}
\usepackage{textcomp}
\usepackage[dvipdfmx]{graphicx}
\usepackage{xcolor}

\graphicspath{{figures/}}

\catcode`\@=11
\def\gta{\ifmmode{\,\mathrel{\mathpalette\@versim>\,}}
    \else{$\,\mathrel{\mathpalette\@versim>}\,$}\fi}
\def\lta{\ifmmode{\,\mathrel{\mathpalette\@versim<\,}}
    \else{$\,\mathrel{\mathpalette\@versim<}\,$}\fi}
\def\@versim#1#2{\lower 2.9truept \vbox{\baselineskip 0pt \lineskip
    0.5truept \ialign{$\m@th#1\hfil##\hfil$\crcr#2\crcr\sim\crcr}}}
\catcode`\@=12  

{\newif\ifnotend
\notendtrue
\def\veclist{ABCDEFGHIJKLMNOPQRSTUVWXYZabcdefghijklmnopqrstuvwxyz.}
\def\top#1#2.{#1}
\def\tail#1#2.{#2.}
\loop\expandafter\xdef\csname bb\expandafter\top\veclist\endcsname%
{{\noexpand\bf\expandafter\top\veclist}}
\edef\veclist{\expandafter\tail\veclist}
\if\veclist.\notendfalse\fi\ifnotend\repeat}

\newcommand{\Uz}{U_0}
\newcommand{\Uu}{U_1}
\newcommand{\Ud}{U_2}

\newcommand{\Vz}{V_0}
\newcommand{\Vu}{V_1}
\newcommand{\Vd}{V_2}

\newcommand{\In}{I_n}
\newcommand{\Iz}{I_0}
\newcommand{\Iu}{I_1}
\newcommand{\Id}{I_2}
\newcommand{\Hn}{H_n}
\newcommand{\RB}{{\cal R}}

\newcommand{\Fell}{{\rm F}}
\newcommand{\Eell}{{\rm E}}
\newcommand{\Pell}{\Pi}
\newcommand{\phiu}{\varphi_1}
\newcommand{\uu}{u_1}
\newcommand{\au}{a_1}
\newcommand{\bu}{b_1}
\newcommand{\sn}{{\rm sn}}
\newcommand{\cn}{{\rm cn}}
\newcommand{\dn}{{\rm dn}}
\newcommand{\am}{{\rm am}}

\newcommand{\Dp}{\Delta_{+}}
\newcommand{\Dm}{\Delta_{-}}
\newcommand{\sphi}{\sin\varphi}
\newcommand{\cphi}{\cos\varphi}

\newcommand{\kp}{k'}

\title[The face-on projection of the Miyamoto \& Nagai disks]
{The face-on projection of the Miyamoto \& Nagai disks}

\author[Luca Ciotti]{Luca Ciotti 
\\ \\
Dept. of Physics and Astronomy, University of Bologna, via Piero Gobetti 93/2, I-40129 Bologna, Italy}

\date{Accepted 2023 August 16, Received 2023 August 08, in original form 2023 July 13}

\begin{document}
\maketitle

\begin{abstract}

  The face-on projected density profile of the Miyamoto \& Nagai disks
  of arbitrary flattening is obtained analytically in terms of
  incomplete elliptic integrals of first and second type, by using two
  complementary approaches, and then checked against the results of
  numerical integration. As computer algebra systems do not seem able
  to obtain the resulting formula in any straightforward way, the
  relevant mathematical steps are provided.  During this study, three
  wrong identities in the Byrd \& Friedman tables of elliptic
  integrals have been identified, and their correct expression is
  given.
    
\end{abstract}  

\section{Introduction}

The Myiamoto \& Nagai disks (1975, hereafter MN, see also Nagai
\& Miyamoto 1976, Binney \& Tremaine 2008, Ciotti 2021) albeit not
very realistic, are among the most used disk models in
numerical/theoretical works for their analytical simplicity, allowing
to express in closed form several properties of interest (see,
e.g. Ciotti \& Pellegrini 1996, hereafter CP96, Smeth et al. 2015): in
particular the MN edge-on projected density distribution can be
written as a surprisingly simple algebraic formula (see, e.g., Satoh
1980, and equation A3 in CP96).  By contrast, the MN face-on
projected density is not found in the literature, and the best known
computer algebra systems seem unable to obtain in any straightforward
way the closed-form solution of the projection integral, which is also
missing from the most used tables of integrals such as Gradstheyn \&
Ryzik (2015, hereafter GR15), Prudnikov et al. (1990, hereafter P90),
Byrd \& Friedman (1971, hereafter BF71).

Of course, for {\it all} practical purposes a numerical evaluation of
the projected density is perfectly suitable (and strongly
recommended), however an inspection of the face-on projection integral
reveals immediately its elliptic nature and so, motivated by
curiosity, and by the desire to add the face-on formula to the already
available edge-on formula, the explicit expression is found and 
  reported here.  For the reasons mentioned above, and in order to
minimize the possibility of errors, all the integrations have been
performed by paper and pencil, intermediate results double checked
with available identities in the tables, and finally verified by
comparison with results of numerical integrations. During the
painstaking process, three wrong integrals have been detected in BF71
(which at the best of our knowledge are not reported in the
published Errata), one of them unfortunately also used in other
identities of the book.

The paper is organized as follows. In Section 2 the projection
integral is reduced to a manageable form with the aid of the Poisson
equation, and the final expression for the face-on projected density
profile is given: readers just interested in the formula can stop
  here. In Sections 3 and 4 the formula is proved in two different
ways. As the involved algebra is quite heavy (the fair copy summing up
to more than 130 handwritten pages), only the main steps of the
process are reported, however all the important information and
technical details are given in the Appendix, togheter with the
derivation of the formulae correcting the three wrong identities
discovered in the tables of BF71.

\section{Projection}

The MN potential reads
\begin{equation}
\phi(R,z)=-{GM\over\sqrt{R^2 + (a+\zeta)^2}},\quad\zeta=\sqrt{b^2+z^2}, 
 \label{eq:MNpot}
\end{equation}
where $a$ and $b$ are two scale-lenghts, and $M$ is the total mass of
the system, so that the disk-like density distribution associated with
eq.~(\ref{eq:MNpot}) via the Poisson equation is
\begin{equation}
 \rho(R,z) = {\triangle\phi\over 4\pi G}={M b^2\over 4\pi}{a R^2+(a+3\zeta)(a+\zeta)^2\over 
   \zeta^3 \left[R^2+(a+\zeta)^2\right]^{5/2}}.
\label{eq:MNrho}
\end{equation}
We recall that for $b=0$ the MN model reduces to the razor-thin
Kuzmin-Toomre disk (Kuzmin 1956, Toomre 1963), while for $a=0$ it
  reduces to the Plummer (1911) sphere. The associated face-on
projected density profile, the focus of the paper, is given by
\begin{equation}
  \Sigma(R) =2\int_0^{\infty}\rho\,dz={1\over 2\pi 
    G}\int_0^{\infty}{\triangle\phi}\,dz,
  \label{eq:projint}
\end{equation}
where the reflection property $\rho(R,-z)=\rho(R,z)$ has been taken
into account.  Due to the nested irrationalities appearing in the
expression of the density $\rho$, the first integral in equation above
looks quite formidable, and in fact symbolic integration is not
performed by the most used computer algebra systems.  Fortunately, the
use of the Poisson equation in the projection integral allows for
considerable preparatory simplification. In fact, the part of the
density associated with the term $\partial^2\phi/\partial z^2$ does
not contribute to the projection, a first simplification of the
problem. Notice that this property is shared by all density
distributions with vanishing force field for $\vert z\vert\to\infty$,
and {\it regular}\footnote{The importance of regularity for this
  simplification, in addition to the vanishing of the force field at
  infinity, is best illustrated by the case of the face-on projection
  of razor-thin disks with finite total mass.} at $z=0$.

After normalization to physical scales $M$ and $b$,
we rewrite eq. (\ref{eq:projint}) as
\begin{equation}
  \Sigma(R) ={M\over 2\pi b^2} I(R),
    \label{eq:projintnor}
\end{equation}
where from now on all the lenghts are intended to be normalized to the
scale $b$, the potential to $G M/b$, and the density to
$M/(4\pi b^3)$. The dimensionless potential is
\begin{equation}
  \phi=-{1\over\sqrt{R^2 + (s+\zeta)^2}},\quad\zeta=\sqrt{1+z^2},\quad
  s={a\over b},
\label{eq:MNpotnor}
\end{equation}
and the dimensionless Poisson equation reads $\rho=\triangle\phi$.
The evaluation of the resulting dimensionless integral
\begin{equation}
I(R) =\int_0^{\infty}{1\over R}{\partial\over\partial R}\left(R{\partial\phi\over\partial R}\right)\,dz,
\label{eq:projintfin}
\end{equation}
is our task, and in the following Sections we give two alternative
derivations of the main result of the paper, namely
\begin{eqnarray}
  I(R)&=&(\Uz+R^2\Vz)\cr\cr
          &+&(\Uu+R^2\Vu)\Fell(\phiu,k)+(\Ud+R^2\Vd)\Eell(\phiu,k).
\label{eq:projfin}
\end{eqnarray}
In the expression above $\Fell$ and $\Eell$ are the incomplete
elliptic integrals of first and second kind\footnote{In Mathematica,
  the argument $k$ in elliptic integrals enters as
  $\Fell(\varphi,k)={\rm EllipticF}[\varphi,k^2]$, etc.}  as given in
eq.~(\ref{eq:ellint}), and
\begin{equation}
\phiu=\arccos{\Dm\over\Dp},\quad k^2={\Dp^2-4\over\Dp^2-\Dm^2},
\end{equation}
the quantities $\Dp(R,s)$ and $\Dm(R,s)$ are defined in
eq.~(\ref{eq:parfin1}), and finally the functions $U_i(\Dp,\Dm)$ and
$V_i(\Dp,\Dm)$ are given in eqs.~(\ref{eq:Un})-(\ref{eq:Vd}).

Before discussing the proof of eq.~(\ref{eq:projfin}), we notice that
two preliminary checks can be obtained by comparison with already
known results, namely 1) $I(R)$ in the spherical limit $s=0$, and 2)
the central value $I(0)$ for arbitrary flattening $s$.  In these two
cases the integral in eq.~(\ref{eq:projint}) becomes elementary and no
elliptic integrals are involved, as discussed in Appendix B and C.  In
the first case, $\Dp=2\sqrt{R^2+1}$ and $\Dm=0$, while only
$U_0=8/\Dp^2$ and $V_0=-32/\Dp^4$ survive in
eqs.~(\ref{eq:Un})-(\ref{eq:Vd}), and the projected density of the
Plummer (1911) sphere is recovered as $I(R)=U_0+R^2V_0$ (e.g., see
equation 13.134 in Ciotti 2021).  In the second case, the verification
that for $R=0$ the function $I(R)$ in eq.~(\ref{eq:projfin}) reduces
to equations (A1)-(A2) in CP96, i.e., that
\begin{equation}
I(0)={2+s^2-3sF(s)\over (1-s^2)^2}, 
\label{eq:Icentral}
\end{equation}  
\begin{eqnarray}
F(s)=\begin{cases}
\displaystyle{{\arccos(s)\over\sqrt{1-s^2}},\quad\quad 0\leq s <1,}\cr 
1, \qquad\qquad\qquad s=1,\cr 
\displaystyle{{{\rm arccosh}(s)\over\sqrt{s^2-1}},\qquad s>1,}
\end{cases}
\label{eq:Fs}
\end{eqnarray}  
is more complicated, as the the elliptic integral decomposition in
eq.~(\ref{eq:projfin}) cannot be directly evaluated at $R=0$, due to
the divergence of the functions $U_i$ and $V_i$ at the
origin. However, as described in Appendix C, it can be shown that the
limit for $R\to 0$ of $I(R)$ in eq.~(\ref{eq:projfin}) recovers
eqs.~(\ref{eq:Icentral})-(\ref{eq:Fs}).

\section{Method I}

From inspection of eq.~(\ref{eq:projintfin}) a first natural idea to
reduce the difficulty of integration is to exchange the
radial derivatives with integration over $z$.  Unfortunately the
exchange is impossibile, as the projection of the potential of any
system of {\it finite} total mass diverges logarithmically, a
consequence of the monopole behavior of $\phi$ at large distances from
the origin: the idea is not to be fully discarded though, and we will
return on it in the next Section. Here we prove eq.~(\ref{eq:projfin})
by brute-force integration of eq.~(\ref{eq:projintfin}), after
evaluation of the radial part of the Laplacian under integral.  From
simple algebra
\begin{equation}
  I(R)=2\Iu(R,\infty)-3R^2\Id(R,\infty),
\label{eq:Imet}  
\end{equation}
where $\Iu(R,\infty)$ and
$\Id(R,\infty)$ are the (finite) limits for $z\to\infty$
of
\begin{eqnarray}
\In (R,z)&=&\int_0^z{dz\over [R^2+(s+\zeta)^2]^{n+1/2}}\cr\cr
  &=&\int_1^{\zeta =\sqrt{1+z^2}}{ [R^2+(s+t)^2]^{-n}\,t\, dt\over\sqrt{(t^2-1)[R^2+(s+t)^2]}},
\label{eq:In}
\end{eqnarray}
with $n=1,2$, respectively, and the last expression is obtained with
the substitution $\zeta=\sqrt{1+z^2}$.  Notice that from
eq.~(\ref{eq:MNpotnor})
\begin{equation}
\Iz(R,z)=-\int_0^z\phi\,dz,
\label{eq:I0z}
\end{equation}
with the expected logarithmic divergence for $z\to\infty$.

Now, even if the Legendre reduction theorem (see Appendix A, and
references therein) guarantees that for integer $n$ the
functions $\In$ can be expressed in general in terms of elliptic
integrals (with the exceptions of the elementary cases corresponding
to $s=0$, and to $R=0$), it should be recalled in practical
applications the explicit factorization of the cubic/quartic
polynomial under the square root often leads to different sub-cases
(depending on the nature and relative positions of the zeroes), and to
cumbersome expressions. Quite remarkably, in the MN case the quartic
under the square root in eq.~(\ref{eq:In}) is already nicely
factorized in two quadratics, one with two real zeroes $\pm 1$, and
the other with complex conjugate roots $-s\pm {\rm i}\,R$. This
factorization is considered in BF71 (equation 260.00, and following),
but, unfortunately, the two integrals $I_1$ and $I_2$ are not
reported, nor they are found in other standard references such as GR15
and P90. Moreover, also the latest releases of the most used computer
algebra systems seem unable to evaluate the definite symbolic
integrals in eq.~(\ref{eq:In}), while in case of indefinite
integration they produce unmanageable expressions involving functions
of complex arguments. For all these reasons here we followed the
approach: 1) we perform the whole integration by paper and pencil,
2) whenever possible we compare intermediate results with available
identities in the Tables, and with numerical integrations performed
with Mathematica NIntegrate and Maple evalf(Int) arbitrary precision
commands, and finally 3) we double check eq.~(\ref{eq:projfin})
against the projected density profile obtained by numerical
integration of the first projection integral in eq.~(\ref{eq:projint})
for different values of the disk flattening $s$.

We found that the most efficient approach to this plan is not to solve
directly the integrals $\In$, but to obtain the closed form expression
of the more general functions $\Hn (y)$ missing from BF71, GR15 and
P90, and then specialize them to the present problem, as described in
Appendix B. In practice eq.~(\ref{eq:projfin}) is proved by first
obtaining $I_1(R,\infty)$ and $I_2(R,\infty)$ respectively from
$H_1(y)$ and $H_2(y)$ in eqs.~(\ref{eq:Hu})-(\ref{eq:Hd}) with $a=1$,
$b=-1$, $\au=R$, $\bu=-s$, $y=\zeta\to\infty$, and then combining them
in eq.~(\ref{eq:Imet}), so that the explicit expressions for the
radial functions $U_i(R)$ and $V_i(R)$ functions in
eqs.~(\ref{eq:parfin1})-(\ref{eq:Vd}) are finally established.  In the
process, we detected three wrong identities in BF71, apparently
missing from the available Errata of the book (Fettis 1972, 1981). The
correct formulae and their proofs are reported in Appendix B.

\section{Method II}

As anticipated in Section 3, the idea behind this second approach is
to minimize the difficulties of integration of
eq.~(\ref{eq:projintfin}), by exchanging the operations of integration
and radial derivatives. Unfortunately, $\Iz(R,z)$ in
eq.~(\ref{eq:I0z}) diverges logarithmically for $z\to\infty$, and so
$I(R)$ cannot be obtained as the radial Laplacian of the
(non-existent) $\Iz(R,\infty)$: however the idea is not to be
dismissed. In fact, as suggested by eq.~(\ref{eq:In}), we first obtain
the two putative functions
\begin{equation}
\Iu(R,z)=-{1\over R}{\partial\Iz(R,z)\over\partial R}, 
\label{eq:dif1I}
\end{equation}
\begin{equation}
  \Id(R,z)=-{1\over 3R}{\partial\Iu(R, z)\over\partial R},
\label{eq:dif2I}
\end{equation}
where the two derivatives are taken at arbitrary but finite $z$, and
then we consider their limit for $z\to\infty$, as required by
eq.~(\ref{eq:Imet}). In practice, we evaluate the radial Laplacian of
$\Iz(R,z)$, and then we consider the limit for $z\to\infty$ of the
resulting expression. That this approach is in fact legitimate can be
shown by comparison of the functions $\Iu(R,z)$ and $\Id(R,z)$
obtained from eqs.~(\ref{eq:dif1I})-(\ref{eq:dif2I}) with the
functions $H_1$ and $H_2$ in eqs.~(\ref{eq:Hu})-(\ref{eq:Hd}),
specialized to the MN case ($a=1$, $b=-1$, $\au=R$, $\bu =-s$,
$y=\zeta=\sqrt{1+z^2}$). From eq.~(\ref{eq:dif1I}) we
  expect that the radial derivative of $\Iz(R,z)$ cancels its
logarithmic divergence for $z\to\infty$, being $\Iu(R,\infty)$ in
  eq.~(\ref{eq:In}) convergent; notice also that, at variance with
eq.~(\ref{eq:dif1I}), the verification of eq.~(\ref{eq:dif2I}) can be
performed directly on $\Id(R,\infty)$, by evaluating the derivative of
$\Iu(R,\infty)$, since this latter function is convergent.

The evaluation of the derivative of $\Iz(R,z)$ is not a trivial task,
as one could have (naively) hoped. The starting point is $H_0(y)$ in
eq.~(\ref{eq:Hz}), specialized to the MN case, i.e.
\begin{eqnarray}
\Iz(R,z) &=&{\Dm/\Dp\over\sqrt{AB}}\times\left[\Pi\left(\phiu,n,k\right) -\Fell(\phiu,k)\right]\cr\cr 
  &+& {1\over 2}\ln{\sqrt{AB}\,\dn(\uu)+\sn(\uu)\over \sqrt{AB}\,\dn(\uu)-\sn(\uu)}, 
\label{eq:projpot}
\end{eqnarray}
where all the quantities appearing in equation above are given in
eqs.~(\ref{eq:parfin1})-(\ref{eq:parfin2}), and the Jacobian functions
are computed from eq.~(\ref{eq:jacf}). A careful analysis reveals that
the logarithmic divergence of $\Iz(R,z)$ for $z\to\infty$ (i.e., for
$\zeta\to\infty$) is due {\it both} to the elliptic integral $\Pi$,
and to the logarithmic term, so that some quite not trivial, exact
cancellation is to be expected after taking the radial derivative of
eq.~(\ref{eq:projpot}): that the cancellation is not trivial can be
realized by inspection of the complicated expressions of the
derivatives of elliptic integrals and Jacobian functions reported in
eqs.~(\ref{eq:djacf})-(\ref{eq:dPn}), and in particular from the fact
that the derivatives of the $\Pi$ function with respect to the
parameters are still expressed in terms of the (divergent) $\Pi$
itself. A simplification of the (very) heavy algebra involved in the
verification of eqs.~(\ref{eq:dif1I})-(\ref{eq:dif2I}) is obtained by
noticing that from eqs.~(\ref{eq:parfin1})-(\ref{eq:parfin2}) the
radial dependence of $\Iz(R,z)$ and of $\Iu(R,z)$ only occurs through
the quantitites $\Dp=A(R)+B(R)$ and $\Dm=A(R)-B(R)$.  In turn, from
eq.~(\ref{eq:parfin1}) $\partial A/\partial R = R/A$ and
$\partial B/\partial R = R/B$, so that for a generic function
$f(\Dp,\Dm)$
\begin{equation}
  {\partial f\over\partial R}
  ={4R\over\Dp^2-\Dm^2}\left(\Dp{\partial f\over\partial\Dp}-\Dm {\partial f\over\partial\Dm}\right).
\label{eq:dfdr}
\end{equation}  
With this approach the equivalence for arbitrary $z$ of $\Iu(R,z)$ in
eq.~(\ref{eq:dif1I}) with the specialization of $H_1(y)$ to the MN
case as required by eq.~(\ref{eq:In}), can be finally established.  A
somewhat simpler verification of eq.~(\ref{eq:dif1I}) in the limit of
large $\vert z\vert$ can be also carried out by considering the
expansion of $\Iz$ for $z\to\infty$ up to the order
${\cal O}(1/\zeta^2)$ with the aid of eq.~(\ref{eq:parfin4}), by
evaluation of the radial derivative of the resulting expression, and
finally by considering the limit for $\zeta\to\infty$. As noticed
above, $\Id(R,\infty)$ can be obtained from eq.~(\ref{eq:dif2I})
evaluated directly on $\Iu(R,\infty)$ by using again the
differentiation rule in eq.~(\ref{eq:dfdr}), obtaining an expression
in perfect agreement with $H_2(\infty)$ specialized to the MN case.

Overall, although the method described in this section is more
  elegant than the brute-force approach of Section 3, it actually
  requires an unexpected amount of work.

\section{Conclusions}

We have shown that the face-on projected density of the Myiamoto \&
Nagai (1975) disk of arbitrary flattening can be expressed in closed
form in terms of incomplete elliptic integrals of first and second
kind, and the explicit formula is provided. The proof (based on two
different but strictly related approaches) proceeds by first reducing
the difficulty of integration thanks to the Poisson equation, and then
by application of the Legendre reduction theorem of elliptic
integrals. The resulting integrals are not evaluated symbolically by
the best known computer algebra systems in any straightforward
way, nor are they given in integral tables such as GR15, P90,
and BF71. For these reasons, all the integrations have been performed
by paper and pencil, intermediate results checked against available
identities in the tables (that have been independently rederived as a
sanity check), and the final formula verified by comparison with
numerical integration of the original projection integral. During this
study three wrong identities in BF71 have been identified and
corrected, and are given together their proof in the Appendix.

We conclude by noticing two consequences of the present
  study. The first is that not only the face-on projected density of
  MN disks can be obtained in closed form, but also their face-on
  projected (self-gravitating) velocity dispersion, given by
\begin{equation}
  \Sigma(R)\sigma_{\rm fo}^{2}(R) =2\int_0^{\infty}\rho\sigma_z^2dz,
\end{equation}
where the integrand is given in eq.~(16) of CP96. An inspection of the
integrand shows that the quantity $\Sigma(R)\sigma_{\rm fo}^{2}(R)$
contains {\it elementary} functions only. However, after division by
the face-on $\Sigma(R)$, we conclude that $\sigma_{\rm fo}^2(R)$ of
the self-gravitating MN disks also involves incomplete elliptic
integrals. Of course, as for $\Sigma(R)$, also for
$\sigma_{\rm fo}^2(R)$, a numerical integration perfectly suffices for
all practical purposes, and for this reason we do not discuss this
problem any further. The second and final comment concerns the Satoh
(1980) disks (e.g., see BT08, Ciotti 2021), a family of models
strictly related to the MN disks, whose potential can be written as
\begin{equation}
\phi(R,z)=-{GM\over\sqrt{R^2 -b^2+ (a+\zeta)^2}},\quad\zeta=\sqrt{b^2+z^2}.
 \label{eq:Satohpot}
\end{equation}
Also for these models the edge-on projected density can be written in
terms of elementary functions, while from the formula above an
analysis similar to that of the MN disks shows that their face-on
projected density contains incomplete elliptic integrals, in a formula
analogous to eq.~(7).

\section*{Acknowledgements}

The Referee, James Binney, and Alberto Parmeggiani of the Department
of Mathematics of Bologna University are thanked for useful comments
and suggestions.

\section*{DATA AVAILABILITY}

No datasets were generated or analysed in support of this research.


\appendix
\onecolumn

\section{Elliptic integrals and elliptic functions}

The literature on elliptic integrals and elliptic functions is
immense, and here we just report the results strictly needed in this
paper.  The elliptic integrals of first, second and third kind in
trigonometric form are given respectively by
\begin{equation}
  \Fell(\varphi,k)=\int_0^{\varphi}{d\theta\over\sqrt{1-k^2\sin^2\theta}},\qquad
  \Eell(\varphi,k)=\int_0^{\varphi}\sqrt{1-k^2\sin^2\theta}d\theta,\qquad
  \Pell(\varphi,n,k)=\int_0^{\varphi}{d\theta\over(1-n\sin^2\theta)\sqrt{1-k^2\sin^2\theta}},
\label{eq:ellint}  
\end{equation}
where $\varphi$ is the {\it argument}, $k$ the {\it modulus},
$\kp=\sqrt{1-k^2}$ is the {\it complementary modulus}, and $n$ the
{\it parameter}: in the present study $0\leq k\leq 1$. A reduction
theorem due to Legendre states that every integral of
$Q[x,\sqrt{P(x)}]$, where $Q(x,y)$ is a generic rational function of
two variables, and $P(x)$ is a polynomial of degree not higher than 4,
can always be reduced to an integral of a rational function of $x$
(and so in principle integrable by elementary methods), and to a
linear combination of $\Fell$, $\Eell$, and $\Pi$ (see, e.g. BF71,
Hancock 1958, Chapter 8). The standard change of variable
$\sin\theta=t$ transforms the elliptic integrals in
eq.~(\ref{eq:ellint}) in the elliptic integrals in algebraic (or
Jacobi) form. For assigned $u$, by inverting $\Fell(\varphi,k)=u$ with
respect to $\varphi$, the {\it elliptic amplitude} $\am(u,k)$ is
obtained, from which the Jacobian elliptic functions remain defined:
\begin{equation}
  u=\Fell[\am(u,k),k], \qquad\varphi=\am(u,k),\qquad\sn(u)=\sin\varphi,\qquad\cn(u)=\cos\varphi,\qquad\dn(u)=\sqrt{1-k^2\sn^2(u)}.
  \label{eq:jacf}
\end{equation}
In particular, $\sn(u)$ is the inverse function of the elliptic
integral of first kind in algebraic form, and as usual in the Jacobian
elliptic functions we do not indicate explicitely the dependence on
$k$.  From the inverse function theorem, one immediately proves that
\begin{equation}
  {\partial\,\am(u,k)\over\partial u}=\dn(u),\qquad 
  {\partial\,\sn(u)\over\partial u}=\cn(u)\dn(u),\qquad 
  {\partial\,\cn(u)\over\partial u}=-\sn(u)\dn(u),\qquad 
  {\partial\,\dn(u)\over\partial u}=-k^2\sn(u)\cn(u), 
\label{eq:djacf}
\end{equation}
while with some more work it can be proved that 
\begin{equation}
  {\partial\,\am(u,k)\over\partial k}={\dn(u)\over k\kp^2}\left[
\kp^2\Fell(\varphi,k)-\Eell(\varphi,k)+k^2{\sn(u)\cn(u)\over\dn(u)}\right],
\label{eq:dampk}
\end{equation}
and the partial derivatives of the Jacobian functions with respect to
the modulus $k$ are obtained from eqs.~(\ref{eq:jacf}) and
(\ref{eq:dampk}), in agreement with BF71 (710.50)-(710.53).  The
derivatives of $\Fell(\varphi,k)$, $\Eell(\varphi,k)$, and
$\Pi(\varphi,n,k)$ with respect to the argument $\varphi$ are
elementary. For the other non trivial identities we have
\begin{equation}
{\partial\Fell(\varphi,k)\over\partial k}={\Eell(\varphi,k)\over k\kp^2}-{\Fell(\varphi,k)\over k} -
{k\sphi\cphi\over\kp^2\sqrt{1-k^2\sin^2\varphi}},\qquad 
{\partial\Eell(\varphi,k)\over\partial k}={\Eell(\varphi,k)-\Fell(\varphi,k)\over k},
\label{eq:dFEk}
\end{equation}  
(e.g., BF71 710.07 and 710.09, or GR15 8.123.1 and 8.123.3), 
\begin{equation}
{\partial\Pi(\varphi,n,k)\over\partial k}={k\,\Eell(\varphi,k)\over\kp^2(k^2-n)}-{k\,\Pi(\varphi,n,k)\over k^2-n} -
{k^3\sphi\cphi\over\kp^2(k^2-n)\sqrt{1-k^2\sin^2\varphi}},
\label{eq:dPk}
\end{equation}  
(BF71 710.12), and
\begin{equation}
{\partial\Pi(\varphi,n,k)\over\partial n}={\Fell(\varphi,k)\over 2n(n-1)}+{\Eell(\varphi,k)\over 2(n-1)(k^2-n)} - {(k^2-n^2)\,\Pi(\varphi,n,k)\over 2n(n-1)(k^2-n)}-
{n\sphi\cphi\sqrt{1-k^2\sin^2\varphi}\over 2(n-1)(k^2-n)(1-n\sin^2\varphi)}, 
\label{eq:dPn}
\end{equation}  
(BF71 733.00).

\section{Three relevant elliptic integrals}

The integrals $I_n$ in Sections 3 and 4 are special cases of the
general identity 260.07 in BF71
\begin{equation}
 H(y)=\int_a^y{\RB(t)\,dt\over\sqrt{(t-a)(t-b)[(t-\bu)^2+\au^2]}}=
  {1\over\sqrt{AB}}\int_0^{\uu}\RB\left[{m\,\cn(u)+p\over\Dp\cn(u)-\Dm}\right]du,\qquad a>b, 
\label{eq:Hn}
\end{equation}
where $\RB(t)$ is a rational function, and 
\begin{equation}
A^2=(a-\bu)^2+\au^2,\quad B^2=(b-\bu)^2+\au^2, \quad\Dp=A+B\geq\Dm=A-B, \quad m=aB+bA,\quad p=aB -bA,
\label{eq:parg1}
\end{equation}
\begin{equation}
k^2={\Dp^2-(a-b)^2\over 4 A B},\quad \kp^2=1-k^2= {(a-b)^2-\Dm^2\over
  4 A B},\quad\am(\uu,k)=\arccos{\Dm y + p\over\Dp y - m}=\phiu,\quad\uu=F(\phiu,k);
\label{eq:parg2}
\end{equation}
finally $4A B =\Dp^2-\Dm^2$, and all the Jacobian elliptic functions
of $\uu$ associated with $H(y)$ can be obtained from
eqs.~(\ref{eq:jacf})-({\ref{eq:parg2}). In the special case $\au=0$
the quartic under the square root in eq.~(\ref{eq:Hn}) reduces to a
quadratic, so that $H(y)$ is expressible in terms of elementary
functions. $H(y)$ is also
elementary when $\bu=0$, $a=-b$, and $\RB(t)$ is an {\it odd} rational function\footnote{A generic
rational function $\RB(t)$ can be always written as the sum of an
even and an odd rational function, with
$\RB(t)=\RB_1(t^2)+t\,\RB_2(t^2)$.}, so that
eq.~(\ref{eq:Hn}) can be integrated with the natural substitution
$x=t^2$. Both these cases are relevant for the discussion at the end
of Section 2 (see also the last paragraph in Appendix C).

The identity in eq.~(\ref{eq:Hn})-BF71 (260.07) can be established
from the change of variable between $t$ and $u$ defined by the
function inside the square parentesis in the last integral in eq.~(\ref{eq:Hn}): after some careful algebra
\begin{equation}
dt={2AB(a-b)\sn(u)\dn(u)\over [\Dp\cn(u)-\Dm]^2}du,\qquad
(t-a)(t-b)={AB(a-b)^2\sn^2(u)\over [\Dp\cn(u)-\Dm]^2},\qquad
(t-\bu)^2+\au^2={(2AB)^2\dn^2(u)\over [\Dp\cn(u)-\Dm]^2},
\label{eq:changv}
\end{equation}
while the new extremes of integration derive immediately from the two
last identities in eq.~(\ref{eq:parg2}).

As apparent from the second integral in eq.~(\ref{eq:In}), in our
problem $\RB(t)$ in eq.~(\ref{eq:Hn}) is
\begin{equation}
\RB(t)={t\over [(t-\bu)^2+\au^2]^n},\qquad n=0,1,2,
\label{eq:RBn}
\end{equation}
so that from the last expression in eq.~(\ref{eq:changv}) we must
consider the integrals
\begin{equation}
\Hn(y)=
  {1\over 4^n(A B)^{2n+1/2}}\int_0^{\uu}{[m\,\cn(u)+p][\Dp\cn(u)-\Dm]^{2n-1}\over\dn^{2n}(u)}du,
\label{eq:Hnn}
\end{equation}
and then specialize them to $a=1$, $b=-1$, $a_1=R$, $b_1=-s$, $y=\zeta=\sqrt{1+z^2}$.

As $H_1$ and $H_2$ are not explicitely given in BF71, GR15, and P90,
and $H_0$ given in BF71 (361.54) is affected by a serious error, here
we derive the expressions of these three functions by direct
integration of eq.~(\ref{eq:Hnn}) with $n=0,1,2$, adopting the same
nomenclature of BF71 for ease of comparison. We have
\begin{eqnarray}
  H_0(y)&=&{1\over\sqrt{AB}}\times\left[{m\over\Dp}\Fell(\phiu,k)+
            {\Dm(\Dp p +\Dm m)\over\Dp (\Dp^2 -\Dm^2)}\Pi\left(\phiu,{\Dp^2\over\Dp^2-\Dm^2},k\right)\right]\cr\cr\cr 
  &+&
           {\Dp p +\Dm m\over (\Dp^2 
      -\Dm^2)(a-b)}\ln{\sqrt{AB}\,\dn(\uu)+(a-b)\sn(\uu)/2\over \sqrt{AB}\,\dn(\uu)-(a-b)\sn(\uu)/2}, 
\label{eq:Hz}
\end{eqnarray}
\begin{eqnarray}
H_1(y)&=&{1\over 4 (AB)^{5/2}}\times\left[{\Dp m\over k^2}\Fell(\phiu,k)-\left({\Dp m\over k^2}+{\Dm p\over\kp^2}\right)\Eell(\phiu,k)+
               (\Dp p-\Dm m) {\sn(\uu)\over\dn(\uu)}\right.\cr\cr\cr 
         &+&\left. 
        k^2\left({\Dp m\over k^2}+{\Dm p\over\kp^2}\right) {\sn(\uu)\cn(\uu)\over\dn(\uu)}\right],
\label{eq:Hu}
\end{eqnarray}
\begin{eqnarray}
H_2(y)&=&{1\over 48 (AB)^{9/2}}\times 
\left\{
\left[{2+k^2\over k^4}\Dp^3 m+{\Dm^3 p\over\kp^2}-3{\Dp\Dm (\Dp p-\Dm m)\over k^2}\right]\Fell(\phiu,k)\right.\cr\cr\cr 
&-&2\left[{1+k^2\over k^4}\Dp^3 m+ {2-k^2\over\kp^4}\Dm^3 p-{3(1-2k^2)\over 2k^2\kp^2}\Dp\Dm (\Dp p-\Dm m)\right]\Eell(\phiu,k)\cr\cr\cr
&+&3\left[\Dp^2(\Dp p - 3\Dm m)+\Dm^2(3\Dp p - \Dm m)\right]{\sn(\uu)\over\dn(\uu)}\cr\cr\cr 
&+&\left[k^2\Dm^2(3\Dp p -\Dm m)-\kp^2\Dp^2(\Dp p - 3\Dm m)\right]{\sn^3(\uu)\over\dn^3(\uu)}\cr\cr\cr 
&+&2\left[{k^2(2-k^2)\over\kp^4}\Dm^3p+{1+k^2\over k^2}\Dm^3m -{3(1-2k^2)\over 2\kp^2}\Dp\Dm(\Dp p-\Dm m)\right]{\sn(\uu)\cn(\uu)\over\dn(\uu)}\cr\cr\cr      
&+&\left.\left[{k^2\over\kp^2}\Dm^3p -{\kp^2\over k^2}\Dp^3 m-3\Dp\Dm (\Dp p -\Dm m)\right]{\sn(\uu)\cn(\uu)\over\dn^3(\uu)}\right\}. 
\label{eq:Hd}
\end{eqnarray}

{\it Proof}: the expression of $H_0$ can be obtained from BF71
(260.03) with $m=1$, followed by (341.02-03), and finally from
(361.54). In the case of MN disks, when $a=1>b=-1$, it is easy to show
that the formula of interest is the third expression in BF71 (361.54):
{\it unfortunately, the given expression is wrong}, as can be seen by
numerical integration, or by differentiation with respect to the
argument. The error was first spotted in the process of recovering
$H_0$ from integration of eq.~(\ref{eq:Hnn}) with $n=0$, as briefly
illustrated. From a partial fraction decomposition of the integrand in
eq.~(\ref{eq:Hnn}) with $n=0$, we arrive at
\begin{equation}
H_0(y)={1\over\sqrt{AB}}\times\left[{m\,\uu\over\Dp}-{\Dp p +\Dm
    m\over\Dp\Dm}\int_0^{\uu}{du\over
    1+\alpha\,\cn(u)}\right],\qquad\alpha=-{\Dp\over\Dm}\leq -1,
\label{eq:Hzproof}
\end{equation}
where the last integral above is BF71 (361.54), and whose correct
expression is obtained in Appendix B1.2. Notice that from
eq.~(\ref{eq:intwrong}) we have $n=\Dp^2/(\Dp^2-\Dm^2)$, so that
$n-k^2= (a-b)^2/(4AB)>0$, and the third case of eq.~(\ref{eq:f1})
applies. Simple algebra then proves eq.~(\ref{eq:Hz}).

The functions $H_1$ and $H_2$, after the expansion of the integrand in
eq.~(\ref{eq:Hnn}), reduce respectively to a linear combination of the integrals
\begin{equation}
H_{1n}=\int_0^{\uu}{\cn^n(u)\over\dn^2(u)}du,\qquad H_{2n}=\int_0^{\uu}{\cn^n(u)\over\dn^4(u)}du.
\label{eq:H1nH2n}
\end{equation}
About the components of the $H_1$ function, $H_{10}$ is obtained from
BF71 (315.02), $H_{11}$ from BF71 (352.51) corrected for a typo as
given in eq.~(\ref{eq:BF35251}) evaluated for $m=1$, or from GR15
(5.137.4).  Finally $H_{12}$ can be obtained from the last of BF71
(355.01) for $n=0$, $m=1$, $p=2$, or from BF71 (320.02).

About the components of the $H_2$ function, $H_{20}$ is obtained from
BF71 (315.04), and $H_{21}$ again from eq.~(\ref{eq:BF35251})
evaluated for $m=2$. $H_{22}$ can be obtained from the last of BF71
(355.01) for $n=0$, $m=2$, $p=2$, and the result is a linear
combination of $H_{10}$ and $H_{20}$.  Analogously, $H_{23}$ can be
obtained from the last of BF71 (355.01) with $n=0$, $m=2$, $p=3$, and
the result is a linear combination of $H_{11}$ and $H_{21}$. Finally,
$H_{24}$ can be obtained from the last of BF71 (355.01) for $n=0$,
$m=2$, $p=4$, or from BF71 (320.04).

\subsection{Three wrong integrals in BF71}

In the process of verification/calculation of the integrals needed in
this work, two typos and a seriously wrong integral were
  discovered in the magnificent book BF71. A search in the available
Errata (Fettis 1972, 1981) indicates that these corrections are almost
surely unknown/unpublished, and so we report them here.

\subsubsection{The integrals BF71 (352.01) and (352.51)}

Two typos appear in the indefinite integrals BF71 (352.01) and
(352.51). For BF71 (352.01) the correct expression reads
\begin{equation}
\int{\sn(u)\over\dn^{2m}(u)}du={1\over\kp^{2m}}\sum_{j=0}^{m-1}{(-1)^{j+1}k^{2j}\over
  2j+1}\binom{m-1}{j}\left[{\cn(u)\over\dn(u)}\right]^{2j+1}.
\label{eq:BF35201}
\end{equation}
{\it Proof}: replace $\sn(u)$ at numerator with the
aid of the third identity in eq.~(\ref{eq:djacf}), write the
resulting $\dn(u)^{2m+1}$ at the denominator in terms of $\cn(u)$, and
reduce the integral to algebraic form by first setting
$t=k\,\cn(u)$, and then $z=t/\sqrt{\kp^2+t^2}$, corresponding to the
substitution $z=k\,\cn(u)/\dn(u)$ in the original integral. For
integer $m=1,2,\ldots$ a binomial expansion proves
eq.~(\ref{eq:BF35201}).

For BF71 (352.51) the correct expression reads
\begin{equation}
\int{\cn(u)\over\dn^{2m}(u)}du=\sum_{j=0}^{m-1}{k^{2j}\over 2j+1}\binom{m-1}{j}\left[{\sn(u)\over\dn(u)}\right]^{2j+1}.
\label{eq:BF35251}
\end{equation}
{\it Proof}: replace $\cn(u)$ at numerator by using
the second identity in eq.~(\ref{eq:djacf}), reduce the
integral to algebraic form by first setting $t=k\,\sn(u)$, and then
$z=t/\sqrt{1+t^2}$, corresponding to the substitution
$z=k\,\sn(u)/\dn(u)$ in the original integral.  For integer
$m=1,2,\ldots$, a binomial expansion finally proves
eq.~(\ref{eq:BF35251}).

\subsubsection{The integral BF71 (361.54)}

The integral is involved in the evaluation of the function $H_0$ in
eq.~(\ref{eq:Hz}). The identity
\begin{equation}
  \int_0^u{dv\over 1+\alpha\, \cn(v)}=
  {\Pi\left(\varphi,n,k\right)-\alpha f_1(u)\over 1-\alpha^2},\qquad
  \alpha^2\neq 1,\qquad n={\alpha^2\over\alpha^2-1},\qquad\varphi=\am(u),
\label{eq:intwrong}
\end{equation}  
reported both in BF71 (41.03) and (361.54) is correct.
Unfortunately, the third case of $f_1$ reported in (361.54) is wrong,
and this error propagates in other identities in BF71 that are
expressed in terms of $f_1$.  The correct expression reads
\begin{equation}
f_1(u)=\int_0^u{\cn(v)\,dv\over 1-n\,\sn^2(v)}=\int_0^{\sn(u)\over\dn(u)}{dz\over 1+(k^2-n)z^2}=
\begin{cases}
  \displaystyle{{1\over\sqrt{k^2-n}}\arctan{\sqrt{k^2-n}\,\sn(u)\over\dn(u)},\qquad\quad\quad\quad n<k^2,}\cr\cr
\displaystyle{{\sn(u)\over\dn(u)}, \qquad\qquad\qquad\qquad\qquad\qquad\qquad\quad\quad n=k^2,}\cr\cr
\displaystyle{{1\over 2\sqrt{n-k^2}}\ln{\dn(u)+\sqrt{n-k^2}\,\sn(u)\over \dn(u) -\sqrt{n-k^2}\,\sn(u)},\qquad\quad n>k^2,}
 \end{cases}
\label{eq:f1}
 \end{equation}
 where from eq.~(\ref{eq:intwrong})
 $\vert n-k^2\vert=(k^2+k'^2\alpha^2)/\vert\alpha^2-1\vert$.  {\it
   Proof}: multiply the numerator and denominator of the integrand in
 eq.~(\ref{eq:intwrong}) by $1-\alpha\,\cn(v)$, and use the identity
 $\cn^2(v)=1-\sn^2(v)$; for $\alpha^2\neq 1$ factor out the quantity
 $1-\alpha^2$ at the denominator, and write the resulting integral as
 the sum of two integrals. The first, from the change of variable
 $\vartheta=\am(v)$ and the use of the first of eq.~(\ref{eq:djacf}),
 or from BF71 (110.04), is immediately recognized as the elliptic
 integral of third kind, while the second is the $f_1$ function, thus
 proving identity in eq.~(\ref{eq:intwrong}). We now focus on the
 $f_1$ function. The $\cn(v)$ function in the integrand of
 eq.~(\ref{eq:f1}) is expressed by using the second of
 eq.~(\ref{eq:djacf}), followed by the substitution $t=\sn(v)$, and
 finally by $z=t/\sqrt{1-k^2t^2}$, corresponding to the single change
 of variable $z=\sn(v)/\dn(v)$. This leads to the second integral in
 eq.~(\ref{eq:f1}), and a last elementary integration proves the three
 cases of $f_1$.

\section{Explicit formulae for the face-on projection of the MN disk}

The functions $I_n$ in Section 3, pertinent to the MN disk face-on
projection, are obtained from the functions $H_n(y)$ in
eq.~(\ref{eq:Hnn}) for $a=1$, $b=-1$, $\au=R$, $\bu=-2$,
$y=\zeta=\sqrt{1+z^2}$, when eqs.~(\ref{eq:parg1})-(\ref{eq:parg2})
reduce to
\begin{equation}
A(R)=\sqrt{R^2+(1+s)^2},\quad B(R)=\sqrt{R^2+(1-s)^2},\quad \Dp=A+B,\quad \Dm=A-B,\quad m=-\Dm,\quad p=\Dp, 
\label{eq:parfin1}
\end{equation}
\begin{equation}
k^2={\Dp^2-4\over 4 A B},\quad k'^2=1-k^2= {4-\Dm^2\over 4 A B},\quad 
n={\Dp^2\over 4 A B},\quad n-k^2={1\over A B},\quad\am(\uu,k)=\arccos{\Dm\zeta + \Dp\over\Dp\zeta +\Dm}=\phiu, 
\label{eq:parfin2}
\end{equation}
and where of course $4AB=\Dp^2-\Dm^2$. Moreover, 
\begin{equation}
  \cn(\uu)= {\Dm\zeta + \Dp\over\Dp\zeta +\Dm},\quad 
  \sn(\uu)={\sqrt{(\Dp^2-\Dm^2)(\zeta^2-1)}\over\Dp\zeta+\Dm},\quad 
  \dn(\uu)=\sqrt{1-{(\Dp^2-4)(\zeta^2-1)\over (\Dp\zeta+\Dm)^2}}, 
\label{eq:parfin3}
\end{equation}
where the asymptotic behavior for $\zeta\to\infty$ at the order
${\cal O}(1/\zeta^2)$ can be easily obtained:
\begin{equation}
  \cn(\uu)\sim {\Dm\over\Dp} +{\Dp^2-\Dm^2\over\Dp^2\zeta},\quad 
  \sn(\uu)\sim{\sqrt{\Dp^2-\Dm^2}\over\Dp}\left(1-{\Dm\over\Dp\zeta}\right),\quad 
  \dn(\uu)\sim {2\over\Dp} +{(\Dp^2-4)\Dm\over 2\Dp^2\zeta}.
\label{eq:parfin4}
\end{equation}
Finally, the following identities used in Section 4 can be established with simple
algebra by using eq.~(\ref{eq:dfdr})
\begin{equation}
{\partial\phiu\over\partial R}=8R {\Dp\Dm\over (\Dp^2-\Dm^2)^2}\sn(\uu),\qquad 
{\partial n\over\partial R}=-16R {\Dp^2\Dm^2\over (\Dp^2-\Dm^2)^3},\qquad
{\partial k\over\partial R}=4R {\kp^2\Dp^2-k^2\Dm^2\over k(\Dp^2-\Dm^2)^2}.
\label{eq:difparam}
\end{equation}

From the general formulae (\ref{eq:Hu})-(\ref{eq:Hd}) we have
\begin{equation}
\Iu(R,\infty)=H_1(\infty)={\Uz+\Uu\Fell(\phiu,k)+\Ud\Eell(\phiu,k)\over 2},\qquad \Id(R,\infty)=H_2(\infty)=-{\Vz+\Vu\Fell(\phiu,k)+\Vd\Eell(\phiu,k)\over 3},
\end{equation}  
to be used in eqs.~(\ref{eq:Imet}) and (\ref{eq:projfin}), and after
some careful algebra we finally obtain
\begin{equation}
  \Uz={32\over (\Dp^2-\Dm^2)(4-\Dm^2)},\quad 
  \Uu=-{16\Dp\Dm\over (\Dp^2-\Dm^2)^{3/2}(\Dp^2-4)},\quad 
  \Ud=-{16\Dp\Dm (\Dp^2+\Dm^2-8)\over (\Dp^2-\Dm^2)^{3/2}(\Dp^2-4)(4-\Dm^2)}. 
  \label{eq:Un}
\end{equation}

\begin{equation}
\Vz=64{\Dm^6+\Dm^4(7\Dp^2-36)-4\Dm^2(5\Dp^2-24)-8\Dp^2(\Dp^2-4)\over (\Dp^2-\Dm^2)^3(\Dp^2-4)(4-\Dm^2)^2}, 
  \label{eq:Vz}
\end{equation}

\begin{equation}
\Vu=-32\Dp\Dm {\Dm^4(\Dp^2-12)+\Dm^2(7\Dp^4-28\Dp^2+64)-8\Dp^2(3\Dp^2-8)\over (\Dp^2-\Dm^2)^{7/2}(\Dp^2-4)^2(4-\Dm^2)}, 
  \label{eq:Vu}
\end{equation}

\begin{equation}
\Vd=-32\Dp\Dm {\Dm^6(\Dp^2-12)+2\Dm^4(7\Dp^4-42\Dp^2+104)+\Dm^2(\Dp^6 -84\Dp^4+352\Dp^2-512)-4\Dp^2(3\Dp^4-52\Dp^2+128)\over (\Dp^2-\Dm^2)^{7/2}(\Dp^2-4)^2(4-\Dm^2)^2}. 
  \label{eq:Vd}
\end{equation}

Concerning the discussion at the end of Section 2 about the central
value of the face-on MN projected density it is important to note that
eqs.~(\ref{eq:Un})-(\ref{eq:Vd}) cannot be evaluated at $R=0$ by
direct substitution, because $\Dp(0)=2$ for $0\leq s\leq 1$, and
$\Dm (0)= 2$ for $s\geq 1$, and so the associated denominators in the
formulae above vanish at the origin. However, it is possible to show
that the limit for $R\to 0$ of the three functions $U_i+R^2V_i$ in
eq.~(\ref{eq:projfin}) exists $\forall s\geq 0$. In particular, from
eq.~(\ref{eq:parfin2}) it follows that at the center $k\to 0$ for
$0\leq s < 1$, and $k\to 1$ for $s>1$.  In the same intervals of $s$,
and for $\zeta\to\infty$ and $R\to 0$, we have $\phiu\to\arccos(s)$
and $\phiu\to\arccos(1/s)$, respectively.  Therefore, for $R\to 0$ the
elliptic integrals in eq.~(\ref{eq:projfin}) can be expressed from
BF71 (111.01) and (111.04) in terms of elementary
functions, in accordance with the first special case $a_1=0$ discussed
after eq.~(B3): the final expression is in perfect agreement with
eqs.~(\ref{eq:Icentral})-(\ref{eq:Fs}), and the case $s=1$ is then
obtained as a limit.
 

\begin{thebibliography}{99}
\bibitem[\protect\citeauthoryear{BT}{2008}]{BT2008}
Binney J., Tremaine S., 2008, Galactic Dynamics, 2nd ed.,
Princeton University Press, Princeton (BT08)

\bibitem[\protect\citeauthoryear{BF}{1971}]{BF1971}
Byrd P.F., Friedman M.N., 1971, Handbook of Elliptic Integrals for Engineers and Scientists. 2nd ed.,
Springer-Verlag (BF71) 

\bibitem[\protect\citeauthoryear{Ciotti}{2021}]{Ciotti2021}
Ciotti L., 2021, Introduction to Stellar Dynamics, Cambridge University Press, Cambridge 

\bibitem[\protect\citeauthoryear{CP}{1996}]{CP1996}
Ciotti L., Pellegrini, S., 1996, MNRAS, 279, 240 (CP96) 

\bibitem[\protect\citeauthoryear{Fettis}{1972}]{Fettis1972}
Fettis H.E., 1972, Math. Comp. 26, 597

\bibitem[\protect\citeauthoryear{Fettis}{1981}]{Fettis1981}
Fettis H.E., 1981, Math. Comp. 36, 315

\bibitem[\protect\citeauthoryear{GR}{2015}]{GR2015}
Gradshteyn I.S., Ryzhik I.M., 2015, Table of Integrals, Series, and Products, 8th ed.,
Zwillinger D. and Moll V. eds., Elsevier (GR15) 

\bibitem[\protect\citeauthoryear{Hancock}{1958}]{Hancock1958}
Hancock H., 1958, Lectures on the Theory of Elliptic Functions, Dover 

\bibitem[\protect\citeauthoryear{Kuzmin}{1956}]{Kuzmin1956}
Kuzmin J. G., 1956, Astron. Zh., 33, 27 

\bibitem[\protect\citeauthoryear{MN}{1975}]{MN1975}
Miyamoto M., Nagai R., 1975, PASJ, 27, 533 (MN) 

\bibitem[\protect\citeauthoryear{NM}{1976}]{MN1976}
Nagai R., Miyamoto M., 1976, PASJ, 28, 1 

\bibitem[\protect\citeauthoryear{Plummer}{1911}]{Plummer1911}
Plummer H.C., 1911, MNRAS, 71, 460 

\bibitem[\protect\citeauthoryear{Prudnikov}{1990}]{Prudnikov1990}
Prudnikov A.P., Brychkov Yu.A., Marichev O.I., 1990, Integrals and Series, Gordon and Breach (P90) 

\bibitem[\protect\citeauthoryear{Satoh}{1980}]{Satoh1980}
Satoh C., 1980, PASJ, 32, 41 

\bibitem[\protect\citeauthoryear{Smet}{2015}]{Smet2015}
Smet C. O., Posacki S., Ciotti L., 2015, MNRAS, 448, 2921  

\bibitem[\protect\citeauthoryear{Toomre}{1963}]{Toomre1963}
Toomre A., 1963, ApJ, 138, 385
  
\end{thebibliography}
\end{document}